\documentclass[showpacs, aps, prb, unsortedaddress,showkeys,longbibliography]{revtex4-1}

\usepackage{amssymb}
\usepackage{latexsym}
\usepackage{amsmath}
\usepackage{graphicx}
\usepackage{float}
\usepackage{xcolor}

\begin{document}

\title{On the Emergence of Negative Effective Density and Modulus in 2-phase Phononic Crystals}

\date{\today}
\author{Amir Ashkan Mokhtari}
\author{Yan Lu}
\author{Ankit Srivastava}
\thanks{Corresponding Author}
\email{asriva13@iit.edu}
\affiliation{Department of Mechanical, Materials, and Aerospace Engineering
Illinois Institute of Technology, Chicago, IL, 60616
USA}

\begin{abstract}
In this paper we report metamaterial properties including negative and singular effective properties for what would traditionally be considered non locally resonant 2-phase phononic unit cells. The negative effective material properties reported here occur well below the homogenization limit and are, therefore, acceptable descriptions of overall behavior. The material property combinations which make this possible were first revealed by a novel level set based topology optimization process which we describe. The optimization process revealed that a 2-phase unit cell in which one of the phases is simultaneously lighter and stiffer than the other results in dynamic behavior which has all the attendant characteristics of a locally resonant composite including negative effective properties far below the homogenization limit. We investigate this further using the Craig-Bampton decomposition and clarify that these properties emerge through an interplay between the fundamental internal modeshape of the unit cell and a rigid body mode. Through explicit numerical calculations on 1-D, 2-phase unit cells, we show that negative effective properties only appear for the specific material property combination mentioned above. Furthermore, we provide a proof which supports this conclusion. The concept is also shown to hold for 2-D unit cells where we show that an appropriately designed hexagonal unit cell made of 2 material phases exhibits negative effective shear modulus and density in an appropriate frequency regime in which it also exhibits negative refraction. As in the 1-D case, we show that the homogenized description in the 2-D case is rigorous as well. An important conclusion of this paper is that the class of unit cells expected to result in negative properties can be expanded beyond the classic unit cell (three-phase unit cells with an explicit locally resonant phase) to include topologically simpler 2-phase unit cells as well. 

\end{abstract}
\keywords{Metamaterial, Phononics, Homogenization, Negative effective Density, Craig-Bampton, Topology Optimization, Double negative metamaterials}

\maketitle

\section{Introduction}
Metamaterials have been an active area of research over roughly the last two decades \cite{srivastava2015elastic}. They are architectured materials which promise material properties which are generally impossible to find in naturally occurring materials \cite{Fang2006, PhysRevE.70.055602,Zhu2014, Liu2011, PhysRevB.93.024302}. As far as acoustic and elastodynamic metamaterials are concerned, the broad objective for their design and existence is to control the flow of sound or stress waves in manners which were not possible before their advent. Some of the wave control objectives include the design of cloaks\cite{1367-2630-8-10-248,1367-2630-9-3-045,Norris2411}, negative refraction and superlenses\cite{Ke2007,Morvan2010,Hu2004}.

\smallskip
\noindent The predominant mechanism for the design and realization of acoustic and elastodynamic metamaterials in literature has been through the introduction of locally resonant phases in the unit cell of a phononic crystals. Under appropriate conditions, the internal resonance of these phases corresponds to singularities and otherwise anomalous behavior in the effective mass, effective density, and/or effective compliance tensors of these composites \cite{milton2007modifications,liu2000locally, huang2009negative}. These locally resonant crystals are predominantly composed of three phases which includes an explicitly locally resonant phase. In some studies, 2-D, 2-phase unit cells composed of an inclusion inside a host material have been noted to display resonance like behavior \cite{PhysRevLett.93.154302} but they have never been associated with negative effective properties. In fact, the presence of resonance is a necessary but not sufficient condition for the emergence of negative effective properties. One of the simplest examples is the case of a 1-D phononic crystal with 2 phases. To our knowledge, this simple unit cell has never been shown to exhibit negative effective properties.

\smallskip
\noindent One of the main contributions of this paper is the elucidation of conditions under which a 1-D, 2-phase phononic crystal exhibits negative effective density and modulus. These conditions were computationally discovered by a multi-phase level set based topology optimization process capable of searching very large spaces in the design domain. Topology optimization has been used in the area of both phononics for the optimization of bandgaps\cite{lu20173,PhysRevE.84.065701}, and in metamaterials for the optimization of metamaterial behavior\cite{Lu2013,KRUSHYNSKA2014179,YANG201689}. We have implemented a comparatively crude form of the level set method geared towards optimizing effective properties. At this point there are different methods available which can be used to calculate the effective properties in metamaterials. Readers are directed towards relevant review papers \cite{srivastava2015elastic,Craster961953} . Here we have implemented a field/ensemble averaging based homogenization scheme for the purpose of optimization \cite{nemat2011overall,nemat2011negative,nemat2011homogenization,srivastava2012overall,srivastava2014limit,willis1983overall,willis2009exact,willis2011effective,SRIDHAR2018104,Muhlestein20160438}. The optimization method as implemented by us is crude but what the method lacks in sophistication, it makes up in decreasing the size of optimization problem and allowing us to deploy global search schemes such as the Basin-Hopping technique \cite{wales1997global}. Using off the shelf codes and algorithms \cite{jones2014scipy}, our optimization algorithm results in interesting optimized unit cells. This includes both locally resonant multi-phase unit cells as well as non-locally resonant 2-phase unit cells exhibiting metamaterial properties. 

\smallskip
\noindent We further explore the dynamics of the 2-phase unit cell suggested by the optimization algorithm through the use of the Craig-Bampton decomposition. This method was introduced to express the dynamic behavior of a structure as a function of its (or its substructures') internal vibrational modes \cite{craig1968}. In the phononics area, it has been used recently to develop highly efficient algorithms for bandstructure and associated calculations \cite{krattiger2014BMS}. In the metamaterials area, it has found natural usage in connecting the effective properties to the internal resonance modes of unit cells \cite{SRIDHAR2018104}. In the present application, the Craig-Brampton decomposition gives invaluable insights into the nature of the interplay between the internal vibration modes of a unit cell and its rigid body motion. For example, the method illuminates that on the second passband, the behavior of the unit cell is dominated by the first vibrational modeshape of the unit cell under either a fixed-fixed or guided boundary conditions -- a conclusion which has been noted earlier using analytical techniques\cite{MEAD19751}. We use the decomposition to highlight the close correspondence between the dynamics of a locally resonant unit cell and a 2-phase unit cell made of an appropriate combination of material properties (suggested by the optimization scheme.) The insights gained from the Craig-Bampton analysis allow us to formally consider the problem of the emergence of negative effective properties in 2-phase phononic crystals. Through explicit numerical simulation and analytical proof we show that appropriately designed 2-phase phononic crystals can exhibit negative effective properties far below the homogenization limit, thus constituting a valid description of overall behavior much like the negative effective properties for locally resonant unit cells.

\smallskip
\noindent In the subsequent sections we first introduce the effective property calculation scheme based on ensemble average of field variables along with two examples, one of which possesses negative effective properties and the other shows only positive values. We then define the level set topology optimization framework in Section \ref{TopOpt} and present the multiphase optimization results as well as a result which shows a two phase unit cell exhibiting negative effective density and elastic modulus. Craig-Bampton decomposition is presented in Section \ref{CBanalysis} followed by its application to the analysis of unit cell dynamics, revealing the central mechanisms which gives rise to negative properties -- an interplay between the first internal vibration mode and rigid body motion. Finally, we apply the essential findings in layered composites to the design of two phase, two dimensional hexagonal unit cell in Section \ref{TwoD}, where we show negative effective properties on the optical branch as well as a simulation of negative refraction in the associated frequency region.

\section{Effective Properties}

\begin{figure}[htp]
\centering
\includegraphics[scale=0.48]{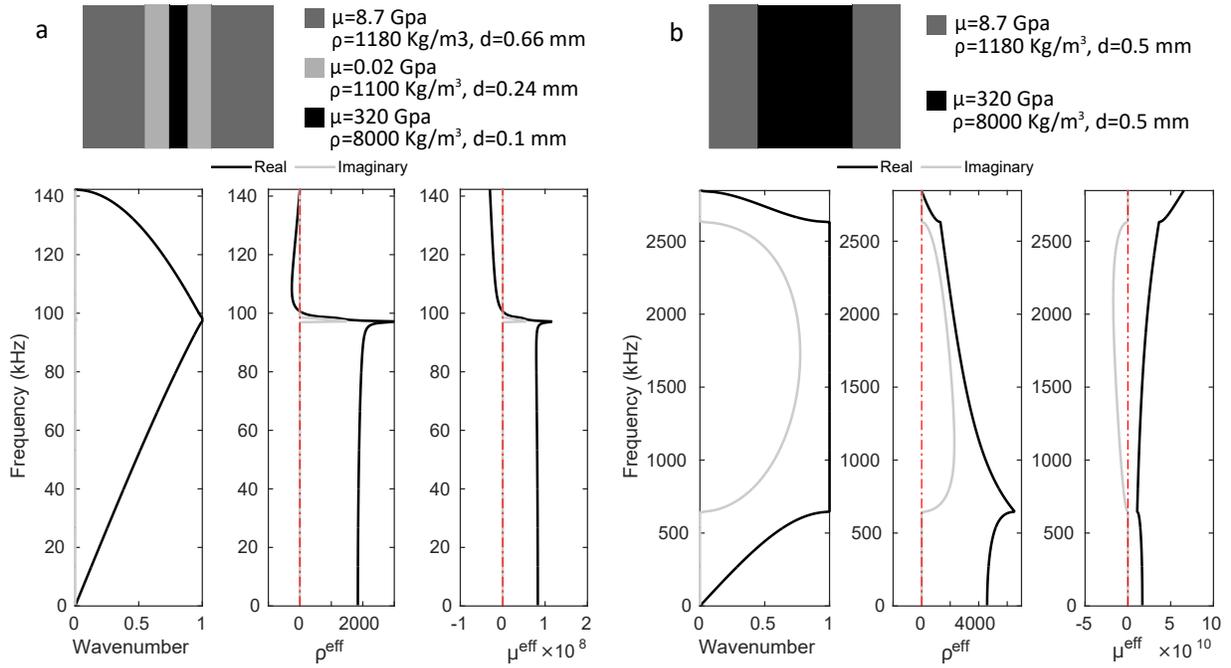}
\caption{Dispersion relations and effective properties of 1-D layered structures. Material properties are excerpted from Ref\cite{nemat2011negative}. Total thickness of each material phase is listed. a. Locally resonant structure. b. conventional 2-phase structure. Zero effective properties are marked by the red dot-dashed lines.}\label{Fig1}
\end{figure}

We consider harmonic waves traveling in a layered composite with a periodic unit cell $\Omega$. Waves are assumed to be normal to the interfaces and in this case the field variables (displacement, velocity, strain, stress) are scalar and take the following Bloch form:
\begin{eqnarray}
\nonumber\displaystyle u(x,t)=U(x)e^{\mathrm{i}(qx-\omega t)};\quad v(x,t)=V(x)e^{\mathrm{i}(qx-\omega t)}\\
\epsilon (x,t)=E(x)e^{\mathrm{i}(qx-\omega t)};\quad \sigma(x,t)=\Sigma(x)e^{\mathrm{i}(qx-\omega t)}
\label{eq:Bloch}
\end{eqnarray}
where functions $U(x)$, $V(x)$, $E(x)$, $\Sigma(x)$, and the momentum $P(x)=\rho(x)V(x)$ are periodic with the periodicity of the unit cell. The dynamic equilibrium and the strain-rate/velocity relations give
\begin{equation}\label{EEquationOfMotion}
\sigma_{,x}+\mathrm{i}\omega p=0;\quad v_{,x}+\mathrm{i}\omega \epsilon=0
\end{equation}
One way of defining the effective properties is through the technique of field averaging \cite{nemat2011homogenization} which is a subset of ensemble averaging \cite{srivastava2015elastic} for periodic composites. Specifically we define:
\begin{eqnarray}
\nonumber\displaystyle \mu^\mathrm{eff}=\langle\Sigma\rangle/\langle E \rangle\;\quad \rho^\mathrm{eff}=\langle P\rangle/\langle V \rangle\\
\langle(\cdot)\rangle=\frac{1}{\Omega}\int_\Omega(\cdot)dx
\label{eq:Effective}
\end{eqnarray}
It can be shown that the above effective properties satisfy both the average equations of motion and the dispersion relation of the composite:
\begin{eqnarray}
\nonumber\displaystyle \langle\Sigma\rangle+\frac{\omega}{q}\langle P\rangle=0;\quad\langle V\rangle+\frac{\omega}{q}\langle E\rangle=0\\
\displaystyle \frac{\mu^\mathrm{eff}}{\rho^\mathrm{eff}}=(\frac{\omega}{q})^2
\label{eq:Average}
\end{eqnarray}
The effective parameters defined by (\ref{eq:Effective}) are real-valued only if the unit cell is symmetric \cite{nemat2011overall,amirkhizi2017homogenization,srivastava2012overall,willis2012construction,srivastava2015elastic}.  For non-symmetric unit cells, the coupling among the field variables renders the effective stiffness, $\mu^\mathrm{eff}$, and mass-density, $\rho^\mathrm{eff}$, complex-valued. Two simple 1-D examples are given in Fig. (\ref{Fig1}). Fig. (\ref{Fig1}-a) shows a locally-resonant 3-phase unit cell which is composed of a stiff and heavy central core coated with a compliant phase, all enclosed in a stiff matrix. The sub-figures below the unit cell show the bandstructure over the first two bands, the corresponding frequency dependent effective density ($\rho^\mathrm{eff}$) and effective modulus ($\mu^\mathrm{eff}$) respectively. The effective properties have been calculated according to Eq. (\ref{eq:Effective}). These figures show that there exists a frequency region on the second band of the band-structure where the effective properties become negative. This behavior does not occur in the case of a regular non-resonant 2-phase unit cell (Fig. \ref{Fig1}-b) where the effective properties are never less than zero in any region over the passbands. The effective properties are always real in the passbands, however, they are complex in the stopbands. Both the real and imaginary parts are plotted in Fig. (\ref{Fig1}).
\smallskip
\noindent Our main motivation in this paper is to elucidate conditions under which 2-phased unit cells can exhibit negative properties. The system which makes this possible was discovered by a topology optimization framework which we describe in the next section.

\section{Topology Optimization Framework}\label{TopOpt}

Topology optimization has evolved rapidly in recent years as a form-finding methodology for structural and materials design\cite{deaton2014survey,sigmund2013topology,cadman2013design,guest2016topology,lu20173}. It seeks to optimize the distribution of material resources across a design domain such that a defined objective function is minimized (or maximized) while the  constraints are satisfied. Typically, finite element methods are used to discretize the design domain and a material relative density $\rho_e$ ranging continuously from $0$ to $1$, is assigned to each element, with  $\rho_e=0$ and $\rho_e=1$ indicating the presence of only material 1 or material 2 in the element, respectively. Intermediate values represent mixtures of the two material phases and are prevented by penalizing their existence, such as through the Solid Isotropic Penalization Method\cite{bendsoe1989optimal, rozvany1991coc}.

\smallskip
\noindent Another category of approaches for topology optimization is based on the level set method. Level set methods were introduced by Osher and Sethian \cite {Osher1988} and they express material boundaries as zero level sets of a function in a higher dimension \cite{ALLAIRE20021125,ALLAIRE2004363,OSHER2001272}. With level set methods, there is no need to explicitly parameterize the moving surfaces and boundaries. For example, considering the topology optimization of a 2-D 1-phase (solid-void) structure for minimum compliance, the level set method proceeds by defining a function $\phi(x,y)$ whose level set $\phi(x,y)=0$ defines the boundary between the solid-void phases. The function $\phi(x,y)=0$ is changed through optimization techniques which implicitly changes the boundaries of the solid-void phase. Here, we use a version of the level set method which is primarily geared towards speed and a potential global search. This is possible to do in our case since we are dealing only with a 1-D case. A description of the technique is presented as follows.
\begin{figure}[htp]
\centering
\includegraphics[scale=.5]{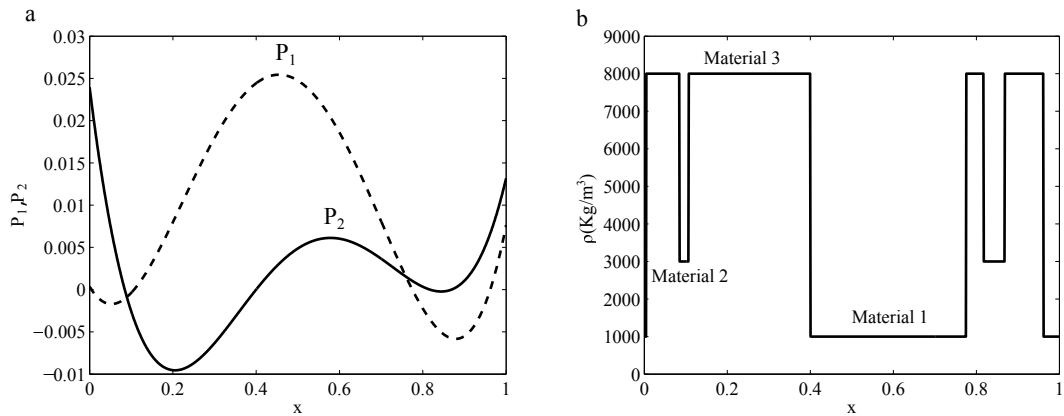}
\caption{(a) Level set functions (b) material distribution corresponding to the level set functions}\label{levels}
\end{figure}
\noindent Consider the topology optimization problem of distributing three different material phases in a 1-D unit cell for achieving some metamaterial objective. Specifically, we want to find a material distribution in this unit cell that leads to negative effective density at a desired point on the irreducible Brillouin zone. The material distribution is governed by the zeros of two polynomials $P_1$ and $P_2$ which are defined over the $x$ axis using the locations of their roots. These two polynomials divide the domain into 3 sub-domains: $\Omega_1\equiv(P_1, P_2>0)$, $\Omega_2\equiv(P_1, P_2<0)$, and $\Omega_3\equiv(P_1 \times P_2<0)$ (Fig \ref{levels}). These three sub-domains are identified as the three different phases of the materials in the design domain. In this technique, the boundaries of these three phases are crisply defined as opposed to the SIMP method of topology optimization where the boundaries are blurry \cite{Bendse1989}. The material distribution is changed by changing the root locations of $P_1$ and $P_2$. In our implementation of the level set method, we have only a few design variables (root locations) for the optimization problem and we can, therefore, use global optimization methods to search very large design-domains. For the optimization algorithm, we use a stochastic technique called Basin Hopping \cite{wales1997global} which attempts to find the global minimum of an objective function. We used the Python Scipy library to run the optimization using Basin Hopping method. The general optimization framework is shown in Fig. (\ref{flowchart}). First, the level set functions are used to initialize the material distribution, and then the zero locations of the level set functions are changed using a canonical Monte Carlo method. In the next step, a local minimization procedure is applied starting from the new configuration to relax the objective function value to the nearby basin (local minimum).  If the new configuration decreases the objective function, it is accepted as the starting point for the next iteration. Otherwise, the old configuration is recovered. The procedure is terminated when the termination condition is satisfied

\begin{figure}[htp]
\centering
\includegraphics[scale=0.5]{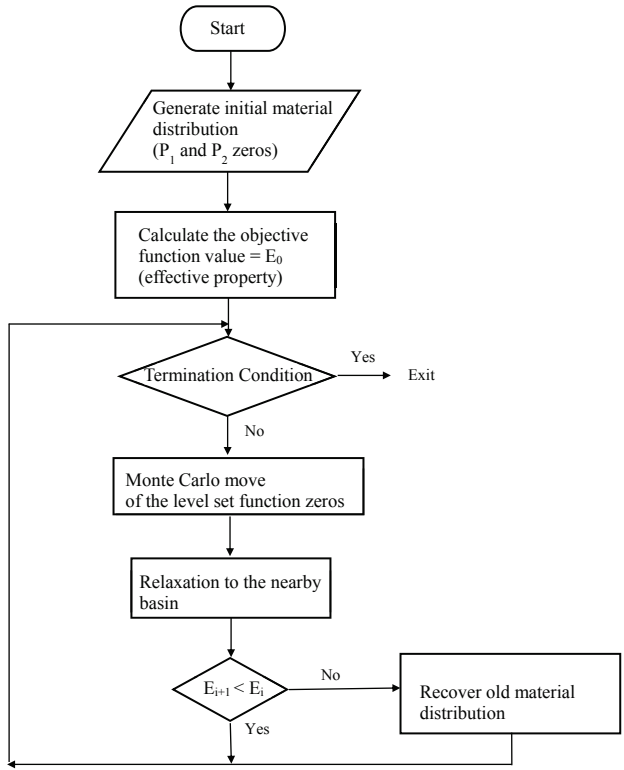}
\caption{General framework of the topology optimization based on the level set and Basin Hopping methods}\label{flowchart}
\end{figure} 

\begin{figure}[htp]
\centering
\includegraphics[scale=0.5]{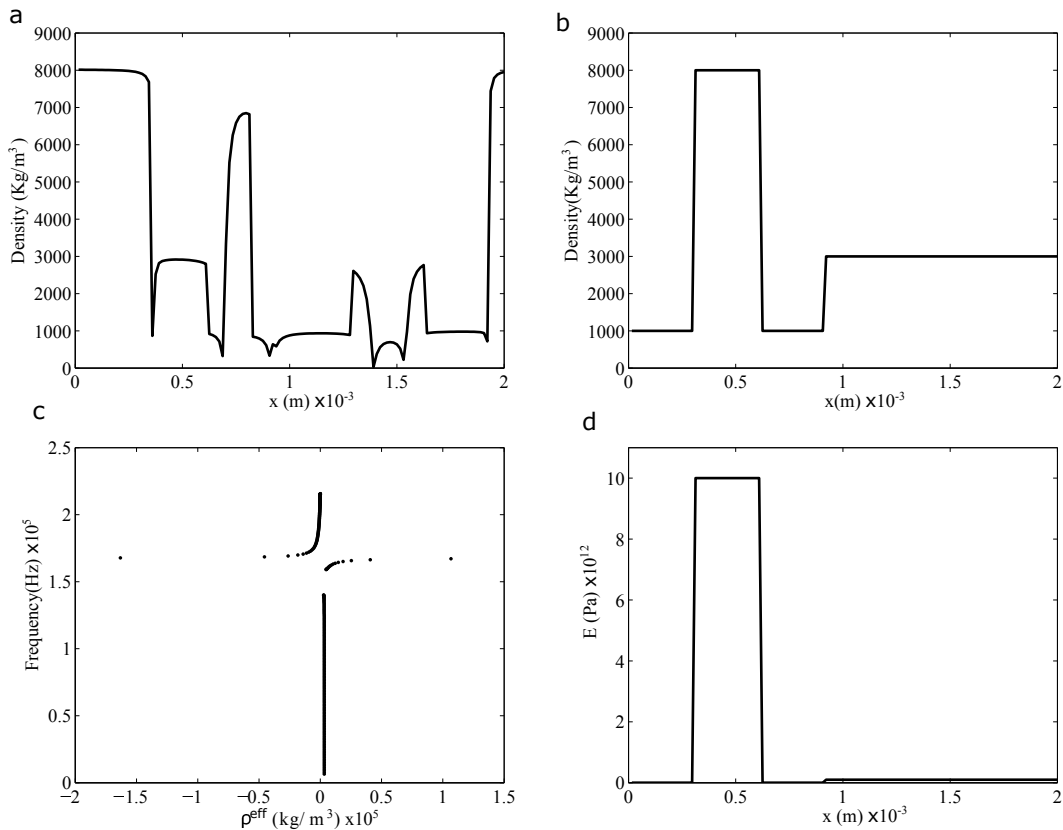}
\caption{(a) Initial random material distribution (b) density distribution in the unit cell (c) effective Density(d) Young modulus distribution}\label{LocalR}
\end{figure}

\noindent  As the first example, the three phases being optimized have material properties which correspond to a traditional locally resonant unit cell. One of the phases $M_1$ is stiff and dense, another phase $M_2$ is compliant and light, and the third phase $M_3$ has stiffness and density values between $M_1,M_2$. For this problem, the objective function to be minimized is the effective density at a desired wavenumber and passband (\ref{eq:Effective}). Fig. (\ref{LocalR}-a) shows the random density distribution (under a relaxation function \cite{Belytschko2003}) over the unit cell taken as the initial condition for the optimization algorithm. Figs. (\ref{LocalR}b,d) are the material property distributions after optimization showing the clean separation of the three phases. When periodicity is considered, this configuration is a traditional local resonance type unit cell with a hard inclusion in the middle ($M_1$), a soft coating layer around it ($M_2$), and all surrounded by a matrix phase ($M_3$). Compared with Fig.( \ref{Fig1}-a), Fig. (\ref{LocalR}-c) shows the effective density calculated over the first two passbands. From Fig. (\ref{LocalR}-c), it can be seen that the effective density goes to a large negative value at a frequency value of around 1600 kHz. The level set optimization algorithm, therefore, naturally finds the characteristic metamaterial unit cell configuration when the optimization objective is the minimization of the effective density. The optimized structure exhibits a negative density region which is expected from an appropriate locally-resonant phononic crystal.
\begin{figure}[htp]
\centering
\includegraphics[scale=.5]{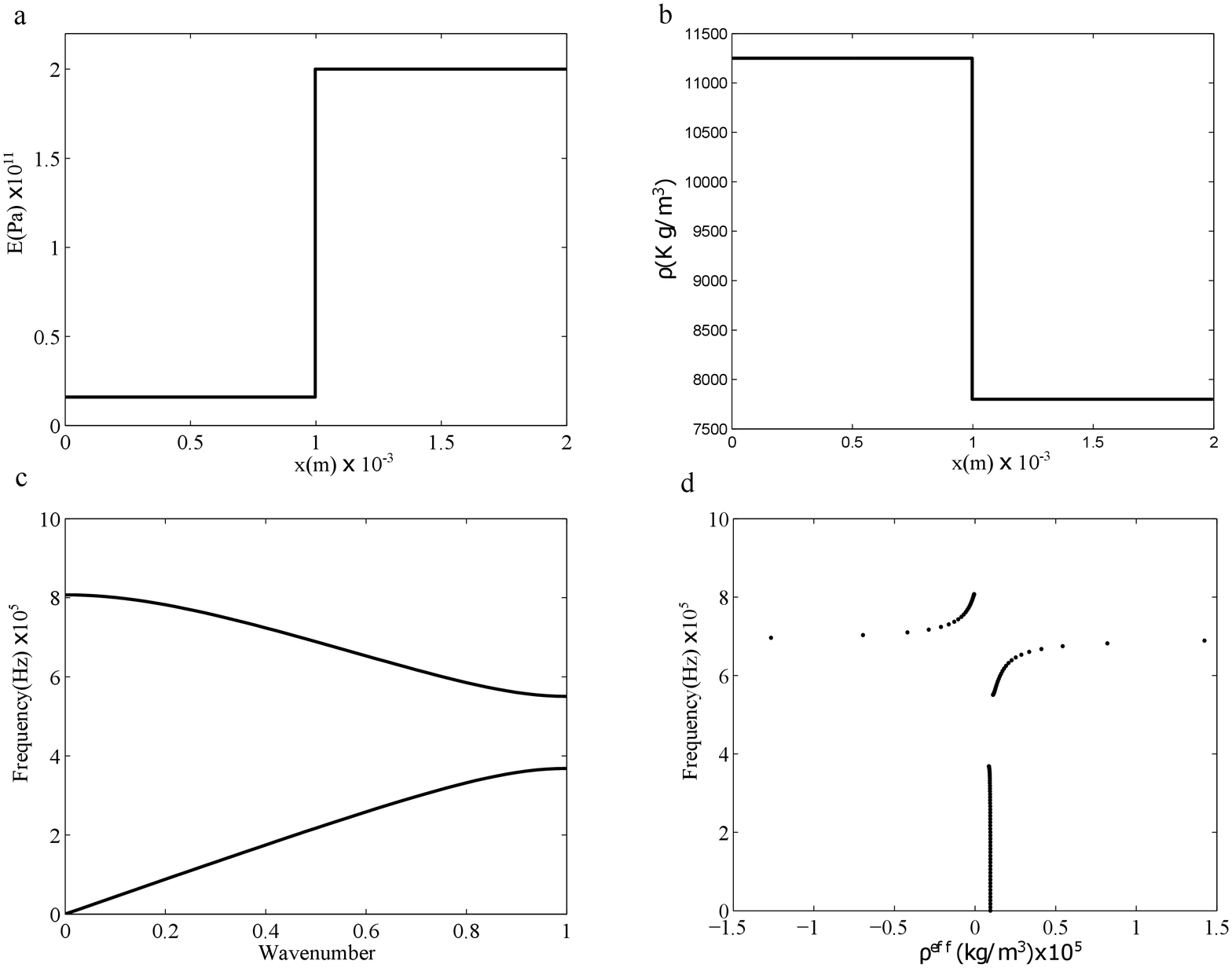}
\caption{(a) Stiffness distribution in the unit cell (b) density distribution in the unit cell (c) band-structure (d) effective density}\label{2phase}
\end{figure}

\noindent The above results are not surprising as the topology optimization framework merely reiterates the unit cell configuration which is traditionally understood to produce negative effective properties in metamaterials. As the next step, in addition to the root locations of the level set polynomials, we also took the material properties (density and Young modulus) in those phases as design variables in the optimization scheme. The optimization objective was to find the material property and geometric distribution combination which would minimize the effective density at a desired wavenumber in a passband. Although a three-phase structure was anticipated as the optimized result in this case, the optimizer, starting with a random distribution of material properties, instead produced a 2-phase structure which exhibited both negative and singular effective density values in some frequency ranges (Fig. \ref{2phase}). The material combination for this structure was different from a conventional 2-phase structure in the following way: in general, there exists a positive correlation between modulus and density of materials \cite{ashby}. In the optimized 2-phased structure produced by the optimizer, one of the material phases was both lighter and stiffer than the other phase ($\rho_1>\rho_2$ and $\mu_1<\mu_2$). The material distribution and effective density plots of the optimized structure are shown in Fig. \ref{2phase}. Figs. (\ref{2phase}a,b) show the stiffness and density distributions in the optimized unit cell, showing that the optimized structure is a 2-phased composite and that it is made of the specific combination of materials that is mentioned above. Figs. (\ref{2phase}c,d) show the bandstructure of this unit cell and the effective density, respectively, underlying the metamaterial nature of this unit cell with the negative and singular values of effective density in some frequency ranges. In the next section, we relate the emergence of negative and singular effective properties to the properties of the internal resonance characteristics of the 1-D unit cell using Craig Bampton reduction method.

\section{Craig-Bampton Analysis}\label{CBanalysis}
Craig-Bampton decomposition \cite{craig1968} is a technique which was first introduced in structural dynamics for the analysis of complex structures in terms of their sub-components.  In this technique, the structure is typically divided in two regions - its boundary and its interior (called the substructure.) The dynamic behavior of the structure is described in terms of a set of generalized coordinates which include constraint coordinates and normal mode coordinates. The constraint modes relate the boundary nodal displacement of the structure to the interior displacements whereas the normal mode coordinates are the vibration modes of the interior with all boundary nodes fixed. This approach, when applied to the present problem, can elucidate the effect of internal resonance on the overall Bloch wave. We summarize the relevant equations for the 1D case below (See Bloch Mode Synthesis \cite{krattiger2014BMS} for more details). Consider the discretized form of the dynamics of a unit cell in a 1D periodic structure:
\begin{equation}
\label{eq:fem}
\mathbf{K}\mathbf{u}+\mathbf{M}\ddot{\mathbf{u}}=\mathbf{f}
\end{equation}
where $\mathbf{K}$ and $\mathbf{M}$ are the nodal stiffness and mass matrices respectively and $\mathbf{u}$, $\mathbf{\ddot{\mathbf{u}}}$ and $\mathbf{f}$ represent the nodal displacements, acceleration and applied forces respectively. The displacement vector can be divided into a set of interior nodes ($\mathbf{u}_I$) and left and right boundary nodes ($\mathbf{u}_B\equiv\{u_{L},u_{R}\}$.) The partitioned form of the Eq \ref{eq:fem} is:
\begin{equation}
\label{eq:fem_partitioned}
\begin{bmatrix}
\mathbf{K}_{II} & \mathbf{K}_{IB}\\
\mathbf{K}_{BI} & \mathbf{K}_{BB}\\
\end{bmatrix}
\begin{bmatrix}
\mathbf{u}_{I}\\
\mathbf{u}_{B}\\
\end{bmatrix}+
\begin{bmatrix}
\mathbf{M}_{II} & \mathbf{M}_{IB}\\
\mathbf{M}_{BI} & \mathbf{M}_{BB}\\
\end{bmatrix}
\begin{bmatrix}
\mathbf{\ddot{u}}_{I}\\
\mathbf{\ddot{u}}_{B}\\
\end{bmatrix}=
\begin{bmatrix}
\mathbf{f}_{I}\\
\mathbf{f}_{B}\\
\end{bmatrix}
\end{equation}
Before applying the Bloch boundary condition on the end nodes $u_{L}$ and $u_{R}$, Craig-Bampton model reduction is used to express the dynamics of the interior nodes in terms of the normal modes and constraint modes as described above. Using the constraint and normal modes, one can replace $\mathbf{u}$ with a new vector $\mathbf{\hat{u}}$ through the following transformation
\begin{eqnarray}
\label{eq:transformation}
\displaystyle \overbrace{\begin{bmatrix}
\mathbf{u}_{I} \\
\mathbf{u}_{B}\\
\end{bmatrix}}^{\mathbf{u}}
=
\overbrace{\begin{bmatrix}
\mathbf{\Phi} & \mathbf{S} \\
\mathbf{0} & \mathbf{I}\\
\end{bmatrix}}^{\mathbf{T}_{1}}
\overbrace{\begin{bmatrix}
\mathbf{\boldsymbol{\eta}} \\
\mathbf{u}_{B}\\
\end{bmatrix}}^{\mathbf{\hat{u}}}\\
\displaystyle
\mathbf{S}=-\mathbf{K}_{II}^{-1}\mathbf{K}_{IB}
\end{eqnarray}
where $\mathbf{\Phi}$ is a matrix whose columns are the vibrational modes of the interior of the unit cell when the boundary nodes are fixed, $\boldsymbol{\eta}$ is a column vector consisting of the amplitude of each normal mode, and $\mathbf{S}$ is a matrix describing the constraint modes. Since $\mathbf{\Phi}$ are vibrational modes corresponding to Dirichlet boundary conditions on the unit cell boundary, we have
\begin{equation}
\label{eq:egnvalue}
(\mathbf{K}_{II}-\omega^{2}\mathbf{M}_{II})\mathbf{\Phi}=\mathbf{0}
\end{equation}
The governing equation (Eq. \ref{eq:fem}) can now be transformed into the new degrees of freedom $\hat{\mathbf{u}}$ which involve the new C-B transformed stiffness and mass matrices:
\begin{eqnarray}
\label{eq:hatK&M}
\mathbf{\hat{K}}=\mathbf{T}^{T}_{1}\mathbf{K}\mathbf{T}\\
\mathbf{\hat{M}}=\mathbf{T}^{T}_{1}\mathbf{M}\mathbf{T}
\end{eqnarray}
At this point in the current problem, the Bloch boundary condition is applied on the left and right boundary nodes $u_{L}$ and $u_{R}$ by the following transformation:
\begin{equation}
\label{eq:BlochBC}
\overbrace{\begin{bmatrix}
\boldsymbol{\eta}\\
u_{L}\\
u_{R}
\end{bmatrix}}^{\hat{\mathbf{u}}}
=
\overbrace{\begin{bmatrix}
\mathbf{I} && 0\\
\mathbf{0} && 1\\
\mathbf{0} && e^{iqh}
\end{bmatrix}}^{\mathbf{T}_{2}}
\overbrace{\begin{bmatrix}
\boldsymbol{\eta}\\
u_{L}
\end{bmatrix}}^{\bar{\mathbf{u}}}
\end{equation}
where $q$ is the wavenumber and $h$ is the unit cell length. This further transforms the stiffness and mass matrices into:
\begin{eqnarray}
\mathbf{\bar{K}}=\mathbf{T}_{2}^{T}\mathbf{\hat{K}}\mathbf{T}_{2}\\
\mathbf{\bar{M}}=\mathbf{T}_{2}^{T}\mathbf{\hat{M}}\mathbf{T}_{2}
\end{eqnarray}
The final form of the eigenvalue problem is as follows:
\begin{equation}
\label{eq:final_egnvalue}
(\mathbf{\bar{K}}-\omega^{2}\mathbf{\bar{M}})\bar{\mathbf{u}}=\mathbf{0}
\end{equation}
The eigenvalues resulting from the solution of the eigenvalue problem give rise to an ordered sequence of radian frequencies $\omega^{(1)}<\omega^{(2)}...$ with corresponding eigenvectors $\bar{\mathbf{u}}^{(1)},\bar{\mathbf{u}}^{(2)},...$. In any eigenvector $\bar{\mathbf{u}}^{(i)}$, the magnitude $\eta^{(i)}_j$ represents the relative contribution of the $j^\mathrm{th}$ vibrational mode in the $i^\mathrm{th}$ eigenvalue-eigenvector pair. With this, the periodic part of the displacement for a given $q$ value and over the $i^\mathrm{th}$ branch, $U^{(i)}(x)$, can be expressed as $U^{(i)}(x)\equiv\{u^{(i)}_L,U^{(i)}_I(x),u^{(i)}_R/e^{iqh}\}$ where 
\begin{equation}
\label{eq:U_I}
U^{(i)}_{I}(x)\equiv U_N(x)+U_B(x)=(\sum_{j=1}^{N} \eta^{(i)}_{j}\Phi_{j}(x) + \mathbf{S}\begin{bmatrix}
u_{L}\\
u_{R}
\end{bmatrix})/e^{iqx}\\
\end{equation}
where $u^{(i)}_{L}$ , $U^{(i)}_{I}(x)$ and $u^{(i)}_{R}$ are concatenated  to form the total mode shape $U^{(i)}(x)$ and $N$ is the number of normal vibration modes considered in Eq. (\ref{eq:transformation}). $U_N(x)$ is the contribution coming purely from the internal modes of vibration and $U_B(x)$ is the contribution coming from constraint mode. As it can be seen from Eq. (\ref{eq:U_I}), $U_I^{(i)}(x)$ is a function of unit cell free vibration normal modes and constraint modes with Bloch boundary condition. When defining effective properties from the solutions above, it is critical that the effective properties, defined in terms of unit cell averages, make physical sense. This essentially translates into the requirement that we should be able to identify a matrix phase which exhibits minimal deformation at the frequency where we are trying to define the effective properties. If this frequency is $\omega^{(2)}$ then this roughly translates into the requirement that $\lambda=2\pi c/\omega^{(2)}$ should be much larger than the size of the matrix phase. In addition, it is required that the wavelength inside each inclusion should be larger than or on the order of the size of the inclusion \cite{PHAM20132125}. 

\paragraph{Locally resonant unit cells} Consider a 3-phase locally resonant unit cell in a symmetric configuration (Fig.\ref{Fig1}-a.) The matrix, inclusion, and the inclusion-coating are indicated by the subscripts $m,i,c$ respectively. If the length of each phase is kept constant, we can follow a concrete set of strategies to induce negative effective properties on the second branch of the band-structure. It has already been shown that for 1-D symmetric unit cells, the frequency limits of the passbands are governed by the natural vibration frequencies of the unit cell in the fixed-fixed and free-free configurations \cite{MEAD19751}. As we will show below, the second branch of locally resonant phononic crystals is primarily influenced by the first vibrational mode of its unit cell in the fixed-fixed configuration which also roughly governs the frequency range of the second branch. If the frequency of this mode is $\omega_0$ then $\omega^{(2)}\approx \omega_0$. Generally $\omega_0$ would change with the unit cell configuration and material properties but it is possible to make it largely independent of large changes in certain properties of the unit cell. Consider the case when the matrix and the inclusion are much stiffer than the coating. In this case, the modeshape of the fundamental vibrational mode in the fixed-fixed configuration of the unit cell displays a behavior where almost all the deformation is in the coating layer. The inclusion moves almost rigidly and there is negligible deformation in the matrix due to its ends being fixed. If we now fix the material and geometric properties of the inclusion and the coating, it is clear that $\omega_0$ will be largely independent of the matrix properties due to the fact that there is almost no deformation in the matrix. In fact, $\omega_0$ will be roughly given by $\sqrt{(\mu_c/l_c)/(\rho_il_i)}$. The wavelength in the matrix at this frequency is $2\pi \sqrt{\mu_m/\rho_m}/\omega_0$ and as long as $2\pi \sqrt{\mu_m/\rho_m}\gg l_m\sqrt{(\mu_c/l_c)/(\rho_il_i)}$, the unit cell can be homogenized. Ignoring the $2\pi$ term, this simplifies into the relation $\beta/\alpha\gg (l_m^2/l_cl_i)$ for homogenization to be applicable where $\beta=\mu_m/\mu_c$ and $\alpha=\rho_m/\rho_i$. If the lengths of the material phases are such that $l_m^2\approx l_cl_i$ then this further simplifies into $\beta/\alpha\gg 1$ or $\beta\gg\alpha$.
\begin{figure}[htp]
\centering
\includegraphics[scale=0.4]{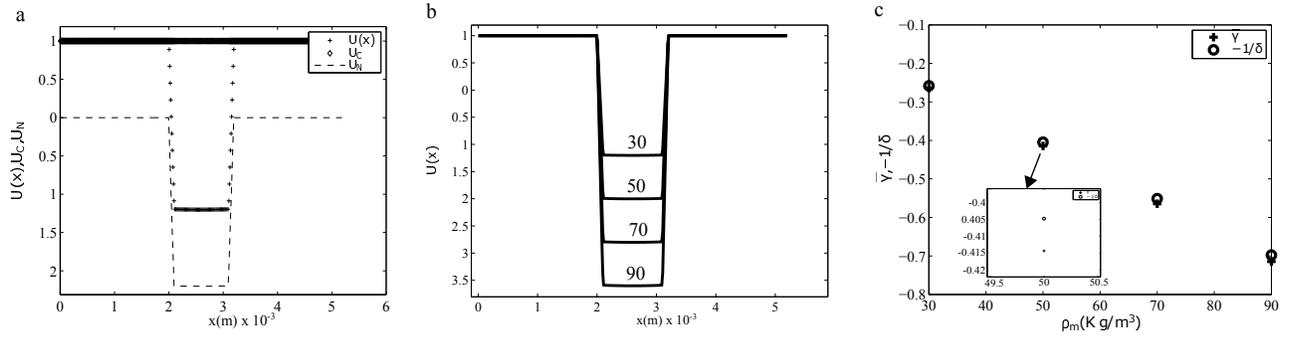}
\caption{(a) Scaled modeshape $U(x)$, $U_{N}$ and $U_{C}$ such that $U_{C}=1$ , (b) modeshapes $U(x)$ plotted for the different values of $\rho_{m}=30, 50, 70, 90 $  (c) $\Bar{\gamma}$ and $-1/\delta$ for $\rho_{m}=30, 50, 70, 90$  }
\label{3-phase_f}
\end{figure}
These ideas can be illustrated by taking an example. Consider the set of cases where $\mu_c=1,\mu_m=10^6$ and $l_m=2,l_c=1/10, l_i=1$. Fig. (\ref{3-phase_f}-a) shows the displacement profile, $U^{(2)}$, of the unit cell for $q\approx 0$ and on the second branch when a specific combination of material properties is assumed ($\mu_i=1000,\rho_i=100,\rho_c=1,\rho_m=30$). Note that for this case $\beta=10^6,\alpha(l_m^2/l_cl_i)=4$ so that homogenization can be performed on the second branch. It also shows the contributions from the internal vibration modes $U_N$ and the boundary constraint mode $U_B$, and the first internal vibration mode of the unit cell $\Phi_1$. There are two points of interest here. First, $U_N$ is almost entirely made up of the first internal mode of vibration of the unit cell or $U_N\propto\Phi_1$, and the boundary constraint mode is almost a rigid body mode or $U_C\propto 1$. Since velocity is directly proportional to displacement, we have $V(x)\approx \gamma \Phi_1(x)+1$ near $q=0$ where $\gamma$ is a scalar and $V$ is the velocity profile corresponding to $U^{(2)}$. The momentum profile is $P=\rho(x)V(x)\approx\gamma \rho(x)\Phi_1(x)+\rho(x)$. Note that in Fig. (\ref{3-phase_f}-a), $\gamma$ is a negative number which elucidates the central mechanism of locally resonant unit cells -- inclusion moving out of phase with the matrix. If $\Phi_1^{i}$ is the velocity of the inclusion in the modeshape $\Phi_1$ then the average of the velocity, $\langle V\rangle$, which appears in effective density equation can be approximated as $\langle V\rangle\approx \gamma l_i\Phi_1^{i}/h+1$, where $h$ is the unit cell length. This is possible since the matrix is nearly stationary and the coating layer is thin. Without any loss of generalization this can be written as $\langle V\rangle\approx \bar{\gamma}+1$ where $\bar{\gamma}=\gamma l_i/h$ and since $\Phi_1(x)$ can always be normalized to set $\Phi_1^{i}=1$. Similarly, the average momentum is $\langle P\rangle\approx \langle\rho\rangle(\bar{\gamma} \delta+1)$ where $\delta=\rho_i/\langle \rho\rangle$ is a positive real number. $\langle V\rangle$ and $\langle P\rangle$ will have opposite signs if $(\bar{\gamma}+1)(\bar{\gamma} \delta+1)<0$. The requirement for this equation to hold is  $-1<\bar{\gamma}<-1/\delta$ (if $\delta>1$ which is true in the present case.) $\delta$ is inversely proportional to $\rho_m$ which means that $-1/\delta$ decreases (increases) with increasing (decreasing) $\rho_m$. Changing $\rho_m$ has no appreciable effect over $\Phi_1$ since the matrix is nearly stationary anyway. However, changing $\rho_m$ serves to change $\bar{\gamma}$. This is illustrated in Fig. (\ref{3-phase_f}-b) where $U^{(2)}$ is plotted for several values of $\rho_m$ ($\rho_i,\rho_c$ are kept constant.) Unsurprisingly, increasing (decreasing) $\rho_m$ serves to decrease (increase) the influence of the rigid body component of motion $U_B$, in turn, decreasing (increasing) $\bar{\gamma}$ in the process. Fig. (\ref{3-phase_f}-c) shows $\bar{\gamma},-1/\delta$ as functions of $\rho_m$ indicating the condition derived above for negative effective properties is indeed satisfied for a range of $\rho_m$. In summary, as long as $\beta\gg\alpha(l_m^2/l_ilc)$, homogenization on the second branch is likely acceptable. Within this constraint the density of the matrix phase can be tuned to modulate the balance between the rigid body component and the first internal mode of vibration component of motion thus giving rise to negative effective density (at least near $q=0$.) Since the effective properties satisfy the dispersion relation of the composite (Eq.\ref{eq:Average}), it is automatically assured that in such regions the effective stiffness will also be negative. Of course, in the above, it is not necessary to assume that the coating layer is thin, however, thin coating layer serves to render the equations in a simple form. 

\paragraph{Two phase unit cells}
\begin{figure}[htp]
\centering
\includegraphics[scale=0.4]{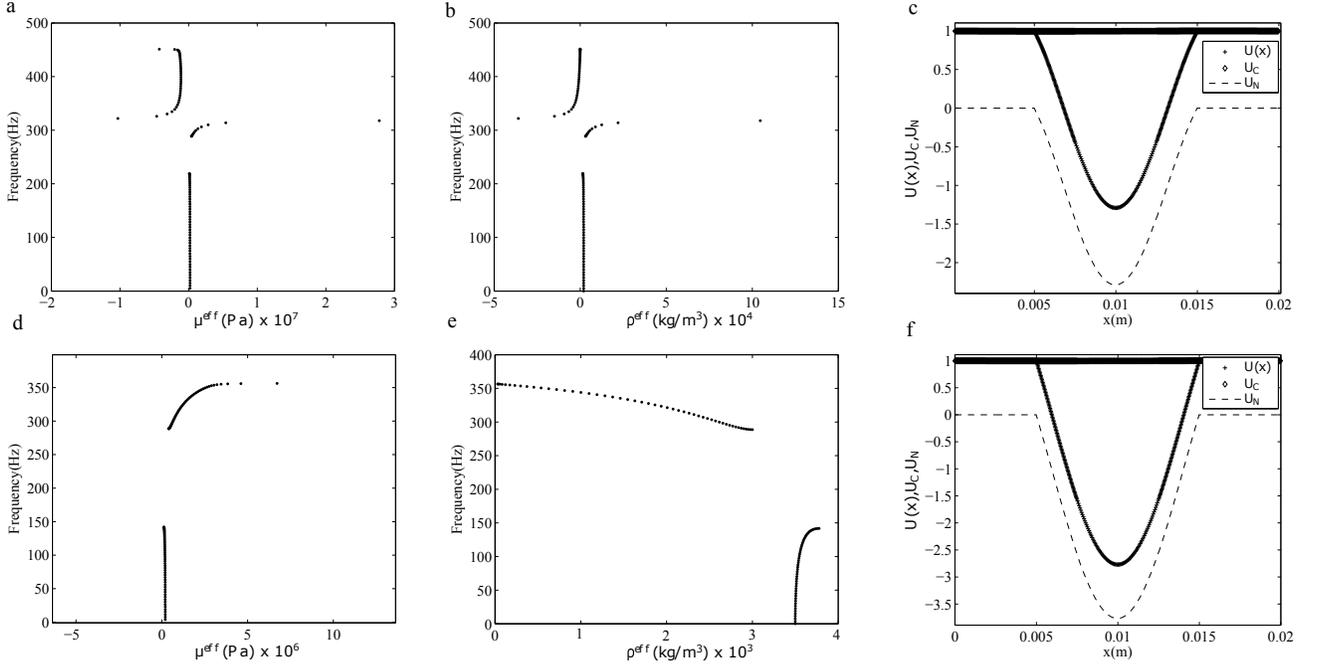}
\caption{(a) Effective stiffness, (b) effective density (c) mode shape $U(x)$, normal mode shape $U_{N}$ and constraint mode $U_{C}$ for the unit cell made of materials with $\rho_{h}=1000$, $\rho_{s}=3000$, $\mu_{h}=1e9$ and $\mu_{s}=1e5$ \\(d) effective stiffness, (e) effective density (f) mode shape $U(x)$, normal mode shape $U_{N}$ and constraint mode $U_{C}$ for the unit cell made of  $\rho_{h}=4000$, $\rho_{s}=3000$, $\mu_{h}=1e9$ and $\mu_{s}=1e5$ }\label{Modes}
\end{figure}
Consider a 2-phase unit cell in the symmetric configuration like Fig. (\ref{Fig1}). In this case, we cannot separate the system into matrix, inclusion and coating layer like the 3-phase case. Instead it has two parts: the soft (with subscript $s$) and hard phase (with subscript $h$). Like the 3-phase case, for a 2-phase unit cell with periodic boundary condition the behavior on the second passband is largely influenced by the first natural vibration modeshape of the unit cell with the fixed-fixed or guided configuration. If the stiffness of one of the phases is large enough compared to the other phase ($\mu_h\gg \mu_s$) then the first free vibration modeshape of the unit cell will comprise of deformation which is concentrated in the soft layer. Again, if the stiffness contrast is high enough, the first natural frequency is determined by the properties of the soft layer which is roughly given as $\omega_{0}=2\pi\sqrt{\mu_{s}/\rho_{s}l_{s}^2}$. This is the result of an eigenvalue problem emerging from a second order homogeneous differential equation in the soft layer with Dirichlet boundary conditions set to zero displacement. The wavelength in the hard layer at this frequency is $\lambda_{h}=2\pi \sqrt{\mu_{h}/\rho_{h}}/\omega_0$ and the relaxed requirement for homogenization is that the wavelength in the matrix should be much larger than the length of the matrix but larger or comparable to the length of the soft phase (inclusion phase.) Note that under the relaxed homogenization requirement, this case is indeed homogenizable. In the vicinity of $\omega_0$, the wavelength in the hard phase is much larger than the length of the phase and, given the modeshape of the first vibrational mode, it is almost twice the length of the soft phase. The nature of the deformation also allows us to consider the harder layer as the matrix. Defining $\beta=\mu_h/\mu_s$ and $\alpha=\rho_{s}/\rho_h$, homogenization is allowed if $\lambda_{h}\gg l_{h}$ or equivalently $\alpha\beta \gg l_{h}^2/l^2_{s}$.

Consider a specific 2-phase unit cell with $h=20 mm$  made of two materials (equal lengths $h_{1}=h_{2}=10 mm$) with $\rho_{h}=1000$, $\rho_{s}=3000$, $\mu_{h}=1e9$ and $\mu_{s}=1e5$, which the second material is heavier and softer than the first one. For the wavenumbers near $q=0$, the normal modeshape $U_{N}$ is almost entirely made up of the first mode of the unit cell natural vibration ($U_N\propto\Phi_1$) and the harder layer is forced into a rigid body motion ($U_C\propto 1$) as depicted in the Fig. (\ref{Modes}-c).  Like the 3-phase case, we have $V(x)\approx \gamma \Phi_1(x)+1$ where $\gamma$ is a scalar and $V$ is the velocity profile corresponding to $U^{(2)}$. The momentum profile is $P=\rho(x)V(x)\approx\gamma \rho(x)\Phi_1(x)+\rho(x)$ as well. If $\Phi_1^{s}$ is the velocity of the soft layer in the modeshape $\Phi_1$ then the average of the velocity, $\langle V\rangle$, which appears in effective density equation can be approximated as $\langle V\rangle\approx \bar{\gamma}+1$ where $\bar{\gamma}=\gamma\int{\Phi_1^{s}}/h$. Similarly the average momentum is $ \langle P\rangle\approx \bar{\gamma}\delta +1$ where $\delta=\rho_s/\langle\rho\rangle$. Following the results from the previous section, effective density would be negative if $-1<\bar{\gamma}<-1/\delta$ (if $\delta>1$) or $-1/\delta<\bar{\gamma}<-1$ (if $\delta<1$.) Of course, since this is a 2-phase system, the requirement for negative properties is $-1<\bar{\gamma}<-1/\delta$ if $\rho_s>\rho_h$ and $-1/\delta<\bar{\gamma}<-1$ if $\rho_s<\rho_h$. The effective property and unit cell deformation results for the present case are graphically shown in Figs. (\ref{Modes}a-c) showing the regions of negative effective properties on the second branch. Figs. (\ref{Modes}d-f) show the corresponding results for another case where the only difference is in the density of the hard phase. Specifically, $\rho_{h}=4000$, $\rho_{s}=3000$, $\mu_{h}=1e9$ and $\mu_{s}=1e5$. Comparing the mode shapes $U(x)$ for both cases (Figs. \ref{Modes}c,f), we note that most of the displacement is in the softer layer in both cases and the harder layer is in a quasi-static state. However, in the first case -- in which the softer phase has also a larger density -- the mode shape is significantly divided into two parts compared to the second case. This is similar to the behavior of a relatively heavier mass vibrating inside a soft material which is characteristic of 3-phase locally resonant unit cells. In contrast, for the second case, in the softer layer most the displacement is in the $U(x)<0$ region. In effect, modulating $\rho_h$ serves to modulate $\bar{\gamma}$ thus giving rise to the potential of negative effective properties. 

\begin{figure}[htp]
\centering
\includegraphics[scale=0.25]{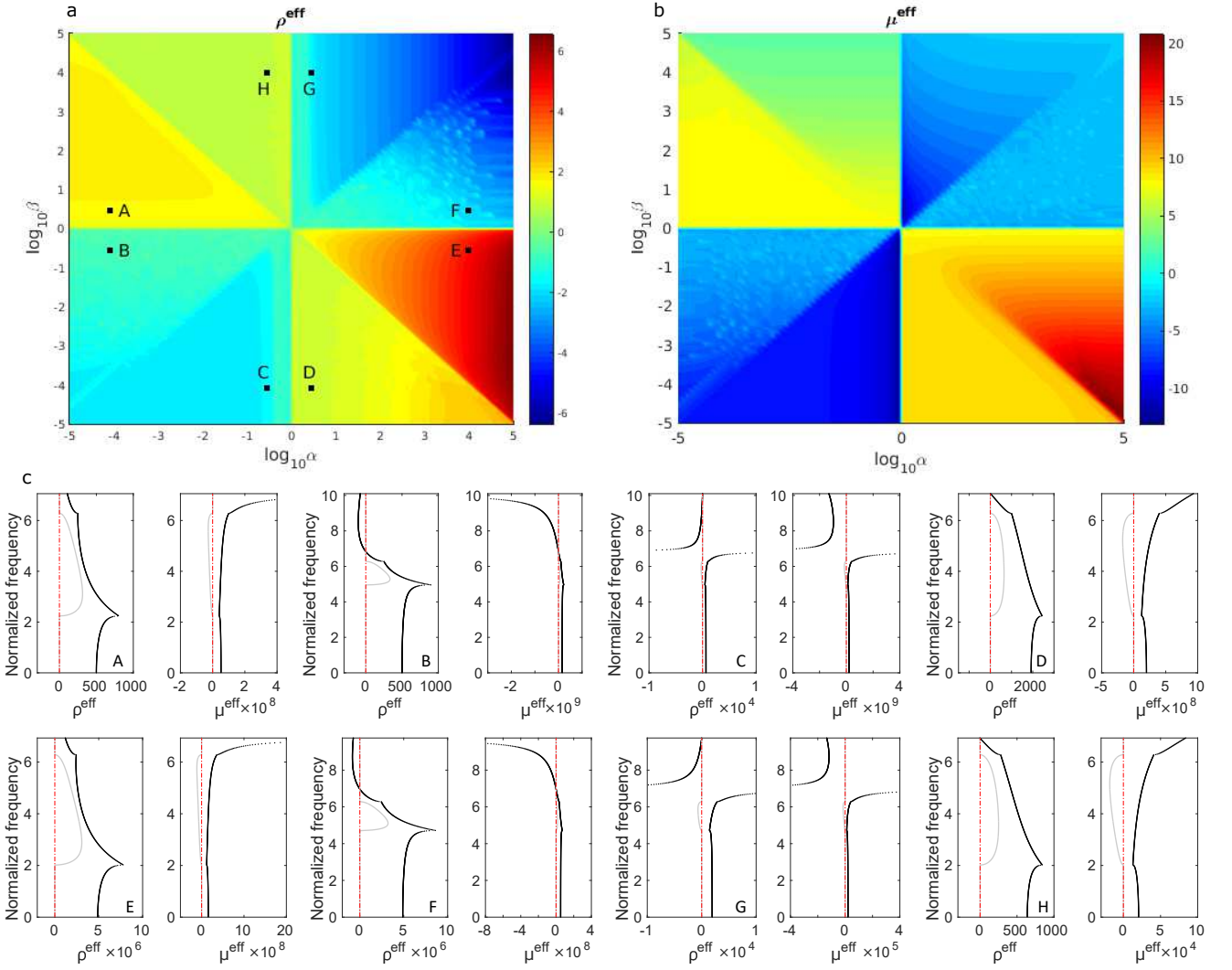}
\caption{(a) Effective density and (b) effective modulus distribution within the frequency range of the second dispersion curve of 2-phase unit cells. Arbitrary units are applied. (c) Real (black) and imaginary (gray) parts of effective properties corresponding to the points marked in (a). }\label{distribution}
\end{figure}

Fig. (\ref{distribution}) shows the effective property distribution ($\rho^\mathrm{eff},\mu^\mathrm{eff}$) as a function of $\alpha,\beta$ for a 2-phase unit cell equal material phase lengths. The material phases $1,2$ have properties $\mu_1,\rho_1,\mu_2,\rho_2$ respectively and $\alpha,\beta$ have been defined as $\alpha=\rho_2/\rho_1$ and $\beta=\mu_1/\mu_2$. Figs. (\ref{distribution}a,b) are log-log plots where the horizontal axis is $\beta=1$ and the vertical axis is $\alpha=1$. $\alpha=\beta$ is the main diagonal. Points $A,B,G,H$ correspond to cases where $\beta\gg\alpha$ and are, thus, situated in homogenizable regions. Points $C,D,E,F$ correspond to cases where $\beta\ll\alpha$ which are also homogenizable when one interchanges the two material phases -- something which is possible in a two phase layered composite. In Figs. (\ref{distribution}a,b), effective properties are evaluated for the second dispersion bands and we only collect the positive and negative values with the smallest absolute values among the discretized frequencies over which the effective properties are calculated. The values are appropriately normalized so that the boundaries can be clearly shown. First it should be noted that Figs. (\ref{distribution}a,b) seem to indicate that negative effective properties are achieved only in 2 of the 4 quadrants. The top right quadrant corresponds to $\alpha,\beta>1$ and the bottom left quadrant corresponds to $\alpha,\beta<1$. In both these cases, one can conclude that negative effective properties are only achieved when one of the phases is simultaneously stiffer and lighter than the other phase. This is explicitly shown for the marked points in Fig. (\ref{distribution}-a) by calculating the effective density and modulus over the first two branches. These points marked are paired and separated by the horizontal and vertical boundaries of the quadrants and they are chosen to demonstrate that by only changing one of the two material property ratios from $<1$ to $>1$, or vice versa, the effective properties can be made to flip signs. These points are also chosen to satisfy $\alpha\beta\gg 1$ or $1/\alpha\beta\gg 1$ so that homogenization of the second branch may be acceptable. The effective property results, shown in Fig. (\ref{distribution}-c), confirm that negative effective properties emerge on the second branch for points $G,F,B,C$ but do not for points $A,H,D,E$.

\paragraph{Required Material Properties for Negative Effective Properties }
\begin{figure}[htp]
\centering
\includegraphics[scale=0.5]{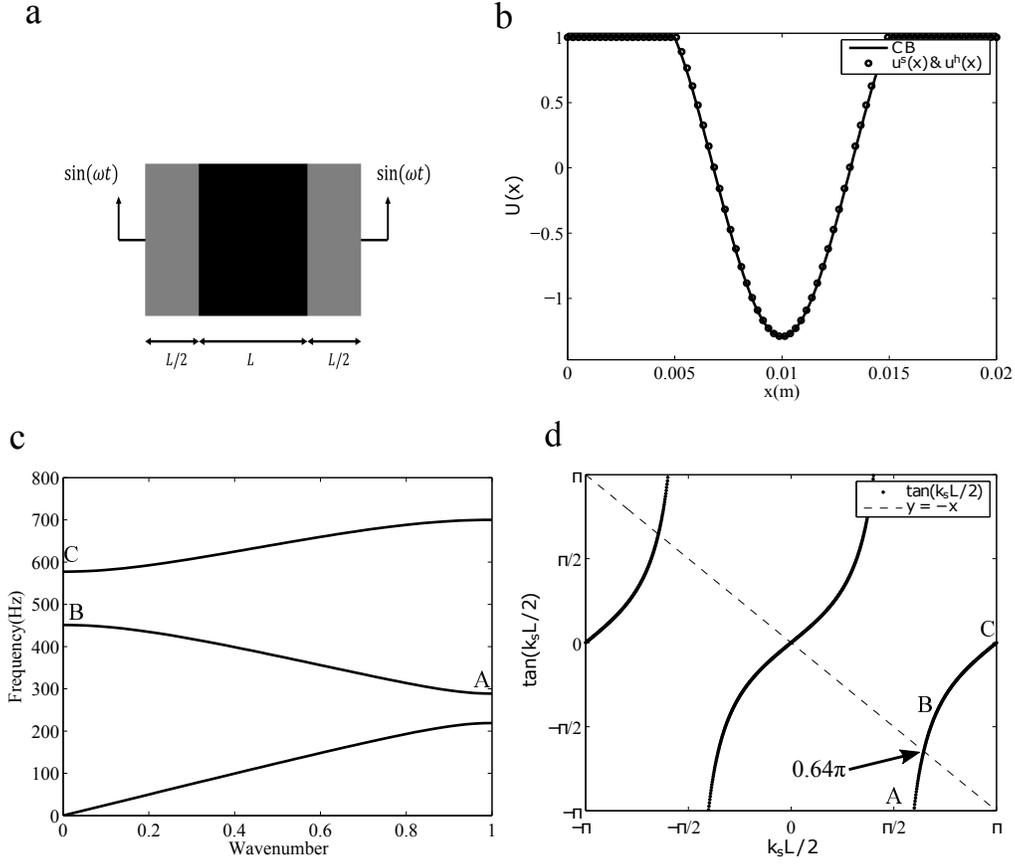}
\caption{(a) Unit cell  with material properties: $\rho_h=1000\,kg/m^3$, $\rho_s=3000 \, kg/m^3$, $\mu_h=1e9 \,Pa$ and $\mu_s=1e5 \, Pa$ (b) Modeshapes calculated using two different methods:  Craig-Bampton and  Eq. (\ref{eq:solution}) (c)  Bandstructure of the unit cell (d) $tan(k_s L/2)$ plot and the corresponding location of the points A,B and C.  }\label{proof}
\end{figure}
The result that negative effective properties only emerge for certain combination of $\alpha,\beta$ can be shown analytically. Assume that the lengths of the phases is the same $l_h=l_s=L$ (Fig. \ref{proof}-a). Also assume that $\beta \gg 1$ and $\alpha >1$. At $q\approx 0$ and on the second branch, we have shown that the total modeshape is a combination of a rigid body mode (because $\beta \gg 1$) and the first vibrational frequency mode ($U(x)\approx\gamma\Phi_1(x)+1$). Let's assume that the displacement profile in just the soft layer is $u^s(x,t)$. The form of the modeshape suggest that $u^s(x,t)$ can be determined by considering the governing equation in the soft layer:
\begin{eqnarray}
\nonumber\displaystyle c_{s}^2 u_{xx}^{s}=u_{tt}^{s}, 
\end{eqnarray}
with the boundary condition:
\begin{equation}
u^{s}(0,t)=u^{s}(L, t)=\sin(\omega t)    
\end{equation}
which imposes Dirichlet boundary conditions corresponding to a rigid body motion $\sin({\omega t})$ to the two ends of the soft layer. In the above, $c_{s}$ is the wave velocity in the soft layer. The solution to this equation is:
\begin{equation}
\label{eq:solution}
u^{s}(x,t)=\left[\frac{1-\cos(\omega L/c_s)}{\sin(\omega L/c_s)}\sin(\omega x/c_s)+\cos(\omega x/c_s)\right]\sin(\omega t)
\end{equation}
The full displacement field in the unit cell which emerges from the above is plotted in Fig. (\ref{proof}-b) showing that it compares very well with the Craig-Bampton solution ($\sin{\omega t}$ term has been ignored in the plot). Having the displacement fields in both of the soft and hard layers, we can find the expression for the effective density as follows:
\begin{equation}
\label{eq:proof_eff}
\rho^\mathrm{eff}=\frac{\rho_h \langle u^h(x)\rangle +\rho_s \langle u^s(x) \rangle}{\langle u^h \rangle + \langle u^s \rangle}=\frac{\rho_h(\alpha\tan(\omega L/2c_s) + \omega L/2c_s)}{\tan(\omega L/2c_s) + \omega L/2c_s}
\end{equation}
where $\alpha=\rho_s/\rho_h$ and $\omega/c_s=k_s$ which is the wavenumber in the soft layer. Fig. (\ref{proof}-c) shows the bandstructure of a typical unit cell. $q=0$ corresponds to point B in the figure and we intend to show that in the vicinity of B, $\rho^\mathrm{eff}$, given in the equation above, is negative if $\alpha$ is greater than 1 and sufficiently large. To show this, we will assume that $\alpha>1$ and show that in the vicinity of point B, the denominator $\tan(\omega L/2c_s) + \omega L/2c_s$ is positive and $\tan(\omega L/2c_s)<0$. This would then automatically mean that $\rho^\mathrm{eff}$ will be negative if $\alpha$ is sufficiently large (from Eq. \ref{eq:proof_eff}).
\smallskip

\noindent The first resonance frequency of the unit cell with the ends fixed coincides with the endpoint of the second pass band (point A in the Fig. \ref{proof}-c) and the second resonance frequency coincides with the beginning of the third passband (point C in Fig. \ref{proof}-c)\cite{MEAD19751}. If we assume that the deformation in the hard layer is negligible, then the frequency at the point A is related to the wavenumber $k_s=\pi/L$ in the soft layer. At this wavenumber the term $\tan(\omega L/2c_s)=\tan(k_s L/2)=\mp\infty$. This point is shown in the Fig. (\ref{proof}-c). Starting from point A on the second passband and moving toward the point C on the $\tan(k_s L/2)$ plot, we are moving from $-\infty$ (for $k_s L/2=\pi/2$) to zero (for $k_s L/2=\pi$). Thus, the tangent term in Eq. (\ref{eq:proof_eff})is negative on the second passband. 

\smallskip

\noindent Now for point B where $q=0$, we have to prove that $\tan(\omega L/2c_s) + \omega L/2c_s >0$. We can write:
\begin{eqnarray}
\label{denom}
\tan(\omega L/2c_s) + \omega L/2c_s =\frac{1-\cos(\omega L/c_s)}{\sin(\omega L/c_s)}+\omega L/2c_s>^?0
\end{eqnarray}
The $?$ mark means that the proof is required for the statement. As explained above, point B is in the range where $\pi<\omega L/c_s<2\pi$, so $\sin(\omega L/c_s)<0$. Thus, we have:
\begin{eqnarray}
\frac{1-\cos(\omega L/c_s)}{\sin(\omega L/c_s)}+\omega L/2c_s>^?0\\
\nonumber \Rightarrow 1-\cos(\omega L/c_s) +\omega L/2c_s\, \sin(\omega L/c_s) <^?0\\
\nonumber \Rightarrow \cos(\omega L/c_s) -\omega L/2c_s\, \sin(\omega L/c_s)>^?1
\end{eqnarray}

\smallskip

\noindent Now we consider the Rytov\cite{rytov1956acoustical} dispersion solution:
\begin{equation}
\label{Rytov}
\cos(\omega L/c_h)\cos(\omega L/c_s)-\Gamma \sin(\omega L/c_h)\sin(\omega L/c_s)=\cos(qh)
\end{equation}
where $\Gamma=(1+\kappa^2)/2\kappa$ and $\kappa=\rho_h c_h/(\rho_s c_s)$. The RHS is equal to 1 in the above at point B. If the hard layer is behaving quasistatically then we can show that at point B $\cos(\omega l_h/c_h)\approx 1$ and $\sin(\omega l_h/c_h) \approx \omega l_h/c_h $. Since the RHS is equal to 1 in the above, we have to show that:
\begin{eqnarray}
\label{Rytov_denom}
\cos(\omega L/c_s) -\omega L/2c_s\, \sin(\omega L/c_s)>\cos(\omega L/c_s)-\Gamma (\omega L/c_h )\sin(\omega L/c_s)
\end{eqnarray}
to complete the proof. From the above we must show:
\begin{eqnarray}
-\omega L/2c_s\, \sin(\omega L/c_s)>-\Gamma (\omega L/c_h )\sin(\omega L/c_s)\\
\nonumber \Rightarrow 1/2c_s> \Gamma (1/c_h)\\
\nonumber \Rightarrow \frac{c_s/c_h+(1/\alpha) \kappa}{\kappa } \approx \frac{\kappa}{\alpha \kappa}<1
\end{eqnarray}
The last part of Eq.(\ref{Rytov_denom}) is true given the assumption that $\alpha>1$. We have, therefore, shown that if $\alpha>1$ then the denominator of Eq. (\ref{eq:proof_eff}) is positive whereas the tangent term is negative. By making $\alpha$ sufficiently large, we can make the numerator negative whereas the denominator is assured to be positive thus giving rise to negative effective density. In fact $ 0.64\pi<k_s L/2 <\pi$ is the range in which the denominator of the Eq. (\ref{eq:proof_eff}) is positive because $\omega L/2c_s > |\tan(\omega L/2c_s)|$. Point B is in this range for the considered combination of $\alpha$ and $\beta$ as proved. This mathematical approach elucidates how the material combination of $\beta=\mu_h/\mu_s>1$ and $\alpha=\rho_s/\rho_h>1$ leads to negative effective properties.

\section{Negative effective properties in 2-D 2-phase unit cells}\label{TwoD}

The essential lessons from the last section also carry on for the design of 2-phase, 2-D unit cells which exhibit negative effective properties. Determination of effective properties in higher dimensions is a much more complicated task than it is in 1-D \citep{srivastava2012overall}. Therefore, we choose an example which is particularly simple but which serves to elucidate the concept. We consider the propagation of antiplane shear waves in a 2-D hexagonal lattice. The displacement field variable of interest is $u_3(x_1,x_2,t)=U_3(x_1,x_2)\exp(i\left[q_1x_1+q_2x_2-\omega t\right])$ where $U_3$ is periodic with the unit cell. The displacement field gives rise to stresses $\sigma_{31},\sigma_{32}$ with unit cell periodic parts $\Sigma_{31},\Sigma_{32}$ respectively. In addition we have velocity ($v_3$), momentum ($p_3$), and strain fields ($\epsilon_{31},\epsilon_{32}$) with periodic parts $V_3,P_3,E_{31},E_{32}$ respectively. The equations of motion are:
\begin{equation}\label{EEquationOfMotion2}
\sigma_{31,1}+\sigma_{32,2}+\mathrm{i}\omega p_3=0;\quad v_{3,1}+\mathrm{i}\omega \epsilon_{31}=0;\quad v_{3,2}+\mathrm{i}\omega \epsilon_{32}=0
\end{equation}
We can define effective properties through field averaging:
\begin{eqnarray}
\nonumber\displaystyle \mu^\mathrm{eff}_{31}=\langle\Sigma_{31}\rangle/\langle E_{31} \rangle\;\quad \displaystyle \mu^\mathrm{eff}_{32}=\langle\Sigma_{32}\rangle/\langle E_{32} \rangle\;\quad \rho^\mathrm{eff}=\langle P_3\rangle/\langle V_3 \rangle\\
\langle(\cdot)\rangle=\frac{1}{\Omega}\int_\Omega(\cdot)dx
\end{eqnarray}
It can be shown that the above effective properties satisfy the average equations of motion:
\begin{eqnarray}
\nonumber q_1\langle\Sigma_{31}\rangle+q_2\langle\Sigma_{32}\rangle+\omega p_3=0;\quad q_1\langle V_{3}\rangle+\omega \langle E_{31}\rangle=0;\quad q_2\langle V_{3}\rangle+\omega \langle E_{32}\rangle=0
\end{eqnarray}
In general, $\mu^\mathrm{eff}_{31}$ and $\mu^\mathrm{eff}_{32}$ will not be the same. Such anisotropy leads to complications as far as determining when a unit cell can be said to exhibit negative effective properties is concerned \cite{srivastava2015causality}. However, if we can find cases where wave propagation is largely isotropic in a certain frequency range, then in that range we can expect $\mu^\mathrm{eff}_{31}=\mu^\mathrm{eff}_{32}=\mu^\mathrm{eff}$. In such cases if $\mu^\mathrm{eff},\rho^\mathrm{eff}<0$ then the composite can unambiguously be said to exhibit negative properties. In such cases, we can also show that the effective properties satisfy the dispersion relation:
\begin{eqnarray}
\displaystyle \frac{\mu^\mathrm{eff}}{\rho^\mathrm{eff}}=\frac{\omega^2}{q_1^2+q_2^2}
\end{eqnarray}

\begin{figure}[htp]
\centering
\includegraphics[scale=0.45]{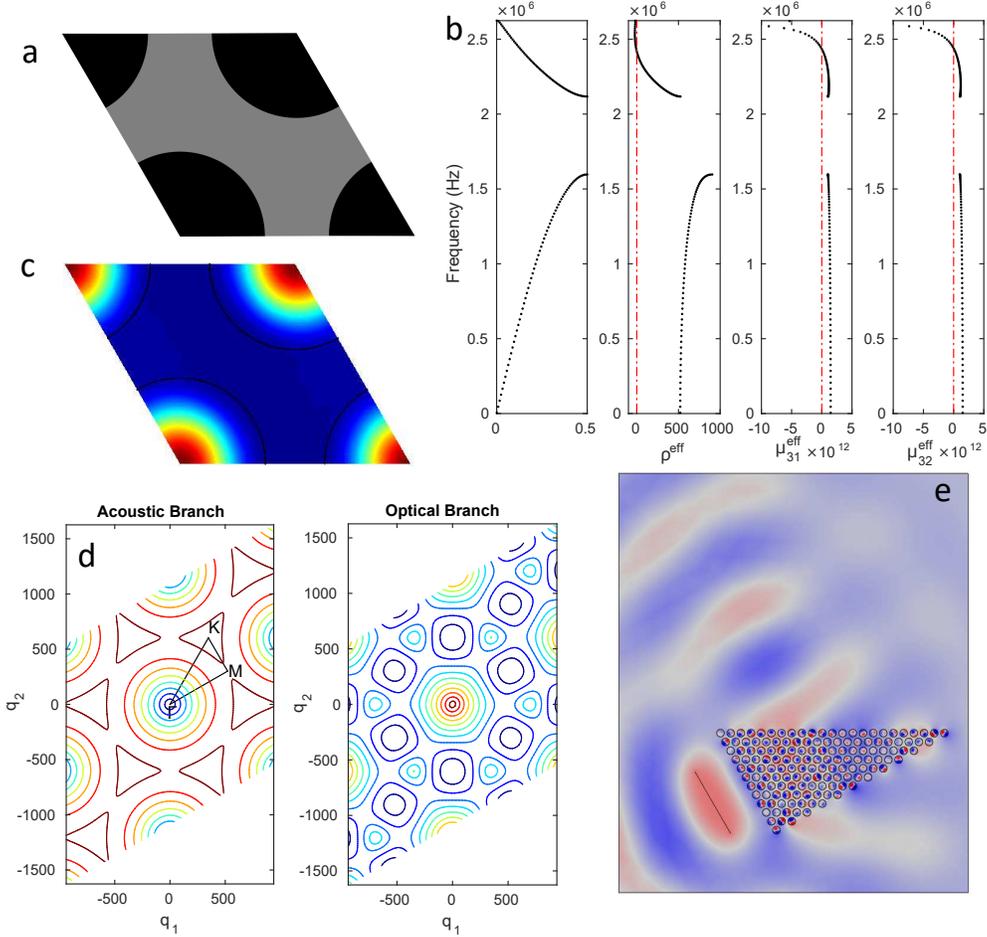}
\caption{(a) Hexagonal unit cell, (b) Band structure and effective properties, (c) Modeshape on the optical branch at 2548$kHz$, (d) Equi-frequency contour, (e) Negative refraction at 2548kHz}\label{Negativeproperties2D}
\end{figure}
For the hexagonal unit cell under consideration, the lattice includes a matrix phase with shear modulus $\mu_m$ and density $\rho_m$. The unit cell contains a symmetric circular inclusion made of a material with shear modulus $\mu_i$ and density $\rho_i$ (Fig. \ref{Negativeproperties2D}-a). The matrix is chosen to be stiff $\mu_m\gg\mu_i$ which ensures that it exhibits minimal deformation up to and around the first vibrational frequency $\omega_0$ of the circular inclusion. Conversely, the circular inclusion is made of a compliant material which ensures that it deforms far easier than the matrix. Furthermore, its density is chosen to be suitably high in order to ensure that $\omega_0$ is small and, thus, ensuring the applicability of homogenization around $\omega_0$. This choice of material property directly follows from the last part of the last section where it was found that material inclusion phases which are simultaneously more compliant and heavier than matrix phases preferentially lead to negative effective properties. The unit cell for a specific material combination satisfying these properties ($\rho_m=40kg/m^3, \mu_m=1000GPa, \rho_i=1000kg/m^3, \mu_i=100GPa$) is shown in Fig. (\ref{Negativeproperties2D}-a) with lattice constant $a=6.023mm$ and filling fraction 0.5. In Fig. (\ref{Negativeproperties2D}-d), we show the equi-frequency contours (EFCs) for both the acoustic and the optical branches for the unit cell. In the contour maps, circular contours represent frequency zones where wave propagation is isotropic. Unsurprisingly, there are circular contours on the acoustic branch signifying regions of quasi-static behavior. More interestingly though, there are circular contours on the optical branch as well. These appear at higher frequencies ($\approx 2500 Hz$) on the optical branch and signify regions where wave propagation is isotropic. Fig. (\ref{Negativeproperties2D}-b), shows the effective properties calculated along the $\Gamma-M$ direction and over the first two branches. It should be noted that the effective properties are negative on the second branch. More importantly though, they are negative in the region around $\approx 2500 Hz$ and in that region $\mu^\mathrm{eff}_{31}$ is ostensibly equal to $\mu^\mathrm{eff}_{32}$. Similar to the 1-D cases in the previous section, in these frequency regions deformation occurs predominantly inside the circular inclusion while the matrix remains in a quasistatic state. This is shown, for example in Fig. (\ref{Negativeproperties2D}-c), where the modeshape of the unit cell at 2548$kHz$ has been plotted. The circular equi-frequency contours on the second branch also predicts negative refraction. Negative refraction is explicitly shown by a frequency domain Finite Element simulation in Fig. (\ref{Negativeproperties2D}-e). Here the incident wave is excited by the lower left line source at 2548$kHz$ and it propagates predominantly perpendicular to the shorter edge of the prism. The prism is made of repeating unit cells of the hexagonal crystal described above. The homogeneous media in which the prism is embedded is the same as the matrix material of the phononic crystal. The corresponding wave number in the homogeneous medium is 101.253rad/m and in the crystal is 67.992rad/m. When the wave reaches the upper interface between the prism and the homogeneous media at 60$^\circ$, the outward negative refraction angle is predicted to be 35.56$^\circ$ from Snell's law. This matches the results from the FE simulation.

\section{Conclusions}
A novel level set topology optimization method has revealed that a 2-phase phononic crystal unit cell in which one of the material phases is simultaneously lighter and stiffer than the other can give rise to negative and singular effective properties on the second branch. The proposed topology optimization method seeks to optimize the material distribution governed by the zeros of level set polynomials. Although in a comparatively crude form, this method allows us to search very large design spaces using global optimizers such as Basin-hopping. Craig-Bampton decomposition is used to gain insight into the emergence of negative effective properties in 2-phase composites. The decomposition elucidates that the second passband is dominated by the fundamental mode of vibration of the soft layer under Dirichlet boundary conditions and it further elucidates the central mechanism of local resonance in layered composites -- the interplay between the deformation in the soft phase with the rigid body motion of the other phase. The method also shows that by tuning the material combinations, the balance between the fundamental mode of vibration and the rigid body component is modulated, thus giving rise to the possibility of negative effective properties. When considering 2-phase unit cells with equal layer thicknesses, we can understand the effect of material contrast ratios on effective properties through explicit calculations. These calculations show that simultaneous negative effective density and stiffness happens when material combination appears in two of the four quadrants where one material phase is stiffer and lighter than the other phase. By slightly changing the contrast ratio from one quadrant to another the effective properties can be made to flip signs. We have provided a mathematical proof which also supports this conclusion. The insights gained from the 1-D case also applies to the 2-D case. We design a 2-D, 2-phase hexagonal unit cell with simple circular inclusions and appropriate material properties and show that it results in negative effective density and shear modulus within an appropriate frequency range. In this range, the wave propagation is nearly isotropic and the matrix behaves quasistatically thus homogenization is acceptable. As in the 1-D case, in this case as well, the deformation occurs predominantly within the soft inclusion and a wave incident at an interface between the crystal and a homogeneous medium suffers negative refraction. 

With the results presented in this paper, we can expand the class of phononic unit cells which exhibit negative effective properties to include simple 2-phase unit cells as well.

\acknowledgments     
 
A.S. acknowledges support from the NSF CAREER grant \#1554033 to the Illinois Institute of
Technology and NSF grant \#1825354 to the Illinois Institute of Technology

%


\end{document}